# Multiple-coil *k*-space interpolation enhances resolution in single-shot spatiotemporal MRI

Gilad Liberman,[1] Eddy Solomon,[1] Michael Lustig[2] and Lucio Frydman[1,*]

[1]*Department of Chemical Physics, Weizmann Institute of Science, Rehovot 76100, Israel and*
[2]*Department of Electrical Engineering and Computer Sciences, University of California, Berkeley, California, USA*

**Keywords:** SPEN, parallel imaging, resolution enhancement, single-scan MRI

**Abstract**

**Purpose:** Spatio-temporal encoding (SPEN) experiments can deliver single-scan MR images without folding complications and with robustness to chemical shift and susceptibility artifacts. It is here shown that further resolution improvements can arise by relying on multiple receivers, to interpolate the sampled data along the low-bandwidth dimension. The ensuing multiple-sensor interpolation is akin to recently introduced SPEN interleaving procedures, albeit without requiring multiple shots.

**Methods:** By casting SPEN's spatial rasterization in *k*-space, it becomes evident that local *k*-data interpolations enabled by multiple receivers are akin to real-space interleaving of SPEN images. The practical implementation of such resolution-enhancing procedure becomes similar to those normally used in SMASH or SENSE, yet relaxing these methods' fold-over constraints.

**Results:** Experiments validating the theoretical expectations were carried out on phantoms and human volunteers on a 3T scanner. The experiments showed the expected resolution enhancement, at no cost in the sequence's complexity. With the addition of multibanding and stimulated echo procedures, 48-slices full-brain coverage could be recorded free from distortions at sub-mm resolution, in 3 sec.

**Conclusion:** Super-resolved SPEN with SENSE (SUSPENSE) achieves the goals of multi-shot SPEN interleaving within one single scan, delivering single-shot sub-mm in-plane resolutions in scanners equipped with suitable multiple sensors.

**INTRODUCTION**

Spatiotemporal encoding (SPEN) presents an alternative to usual time-domain or *k*-space approaches, which can deliver multi-dimensional magnetic resonance (NMR) spectra or images (MRI) in a single scan (1-11). SPEN relies on a linear excitation or inversion of the spins as a function of time. This imposes linear and quadratic evolution phases on the spin-packets throughout the field of view (FOV) (12-14); the linear dephasing is the basis of single-scan 2D NMR spectroscopy (5,15-17), whereas the quadratic one opens an alternative route to single-scan 2D MRI (1-11). SPEN differs from either conventional 2D NMR or MRI, by the fact that the spectral or spatial information that it delivers arises as direct observable in the time-/*k*-domain; *i.e.*, without a need for Fourier transforming the acquired data. In the imaging case this can be understood by considering the consequences of applying a chirped inversion pulse of duration $T_e$ and bandwidth *BW*, while under the action of an encoding gradient $G_e$. Such pulse will impart a parabolic phase on the spins, $\varphi_e(y) = \alpha_e y^2 + \beta_e y + c$, where *y* is assumed to be the encoding gradient's axis. This parabolic profile means that the collected signal will be dominated by spins located at the stationary point fulfilling $[\partial\varphi/\partial y] = 0$; *i.e.* from regions where the spins' accrued phases change slowly (18). While this condition will be initially fulfilled by $y_o = -\beta_e/2\alpha_e$, this stationary point can be displaced by application of additional gradients. SPEN MRI time-domain signals are thus acquired while under the action of an acquisition gradient $G_a$, which moves the stationary point throughout the FOV, rasterizing the NMR profile being sought over an acquisition time $T_a$. This approach to scanning 1D axes can be extended into single-shot 2D MRI experiments by executing it in a so-called "hybrid SPEN" fashion, whereby an orthogonal readout (RO) domain is imaged in a usual *k*-space fashion by the incorporation of a second, rapidly oscillating gradient. The ensuing sequence carries an evident resemblance to Spin-Echo Echo Planar Imaging (SE-EPI, Figure 1), but provides the possibility of imaging the latter's low-bandwidth domain without the limitations of Fourier sampling. This enables the use of stronger gradients than in EPI-based counterparts; it also opens the possibility to perform a "full-refocusing" whereby spin-packets echo continuously throughout the course of $T_a$, rather than at a single instant as in a normal SE (2,19,20). For this full-refocusing condition to hold the acquisition gradients and times have to fulfill $T_e = \frac{T_a}{2}$ and $|G_e| = |G_a|$ for an inversion-based encoding or $T_e = T_a$ when relying on a chirped excitation –this difference stemming from the additional factor of two introduced in the phase by a swept inversion pulse over an excitation counterpart of equal bandwidth *BW*. These possibilities of using stronger gradients and of refocusing $T_2^*$ effects throughout the acquisition, have been shown to greatly aid in the performance of single-scan MRI experiments at high fields, or when targeting challenging areas in animal or human anatomies (6-8,21-25).

Despite these advantages, SPEN faces a number of sensitivity vs resolution challenges. SPEN's resolution is in principle given by $\alpha_e$, that is, by the steepness of its parabolic phase. Imposing a tight parabola will achieve high spatial resolution yet at the expense of sensitivity, an onerous cost given SPEN's non-Fourier nature. Loosening the parabola sacrifices resolution but reinstates an EPI-like sensitivity, while simultaneously lessening power deposition (SAR) requirements (26). Moreover, it has been shown that the use of super-resolution (SR) and of other post-processing algorithms (4,7-10,22) can make up for these resolution losses, provided that the sampling occurring along the SPEN axis is sufficiently dense. In single-shot 2D SPEN (Fig. 1B), however, instances will often arise where the resolution and FOV conditions desired along readout, will not allow a sufficiently dense sampling along the SPEN acquisition domain. Recently, Schmidt *et al* described a way of alleviating this, based on the acquisition of multiple interleaved scans (27). Unlike interleaving in EPI, where small sub-dwell $\Delta k$ "blips" increase the sampling density so as to faithfully cover the targeted low-bandwidth FOV and thus avoid folding (28,29), SPEN interleaving aims to improve the spatial resolution by sampling regions *in space* that would otherwise be skipped by the sampling parabola (Fig. 1, lower panels). This presents multiple advantages vis-à-vis EPI's interleaving, including the delivery of full-FOV images for every interleaved scan, and thereby the option of carrying out the procedure in a self-referenced, image-based fashion (27). Still, as hitherto described, both interleaved EPI and SPEN share the common need for performing multiple scans for improving the images being sought –thus forfeiting the original, single-shot nature of these experiments.

The present study explores the possibility of achieving an oversampling that is identical to that afforded by interleaved SPEN, while confining the experiment to a single scan. This possibility arises from the use of multiple receiving coils. Parallel imaging is widely used in MRI as a way of enlarging the targeted FOV, without complying with the full criteria imposed by Nyquist sampling (28-31). Parallel acquisition methods have also been employed in SPEN to enlarge the FOV while shortening the acquisition time, by using multi-band swept pulses targeting regions that are associated to different receiving coils (3,32). It is interesting to note the differences between these two approaches to exploit the availability of multiple receivers, as these stress the distinct principles which these acquisitions exploit to deliver their images. In SiMultaneous Acquisition of Spatial Harmonics (SMASH), SENSitivity Encoding (SENSE) or GeneRalized Autocalibrating Partially Parallel Acquisition (GRAPPA), the sensitivity of different coils to different regions in space is employed in order to effectively fill skipped *k*-points, and thereby avoid undesired fold-overs (Fig. 1A, lower panel); in SPEN, the availability of coil-resolvable regions in space has been used to expand the FOV by simultaneously encoding multiple adjacent volumes using multiband chirp pulses, and unraveling these by independent sensors. There is, however, a hitherto unexplored option to improve the latter experiment by parallel receiving, and which stems from viewing the SPEN acquisition as a sampling occurring not just in real but also in *k*-space. Cast in that scenario, the resolution enhancement arising upon implementing SPEN image interleaving, can also be visualized as involving the collection of additional scans filling in-between samples in a *k*-space grid. Unlike what happens in regular MRI these *k*-blips will only contain contributions that are spatially limited by the parabolic encoding. Due to this local encoding nature, filling up these "missing" points does not result in a global image unfolding; instead, it provides a potential resolution enhancement that is entirely analogous to that arising by the SPEN interleaving procedure. One can thus envision relying on similar algorithms as used in SMASH or SENSE, in order to extract this additional information; SPEN's localized nature, however, implies that these reconstructions will be much less demanding than their regular *k*-sampling counterparts. The principles of this SPEN-oriented reconstruction algorithm are described in the following section based on more familiar SMASH and SENSE arguments; we then present phantom and *in vivo* results collected on a clinical head setup that illustrate the ensuing super-resolved SPEN with SENSE (SUSPENSE) resolution enhancement, including the methodology's potential to deliver quality single-shot brain diffusion maps at sub-mm resolutions. Furthermore, a number of additional pulse schemes are incorporated into SUSPENSE, that enable it to achieve these resolutions over FOVs, SARs and repetition times (TR) that are equal to state-of-the-art EPI values –while benefiting from SPEN's additional robustness to field heterogeneities. An analysis of SUSPENSE's sensitivity compromises on the basis of g-factor mapping arguments, is also presented.

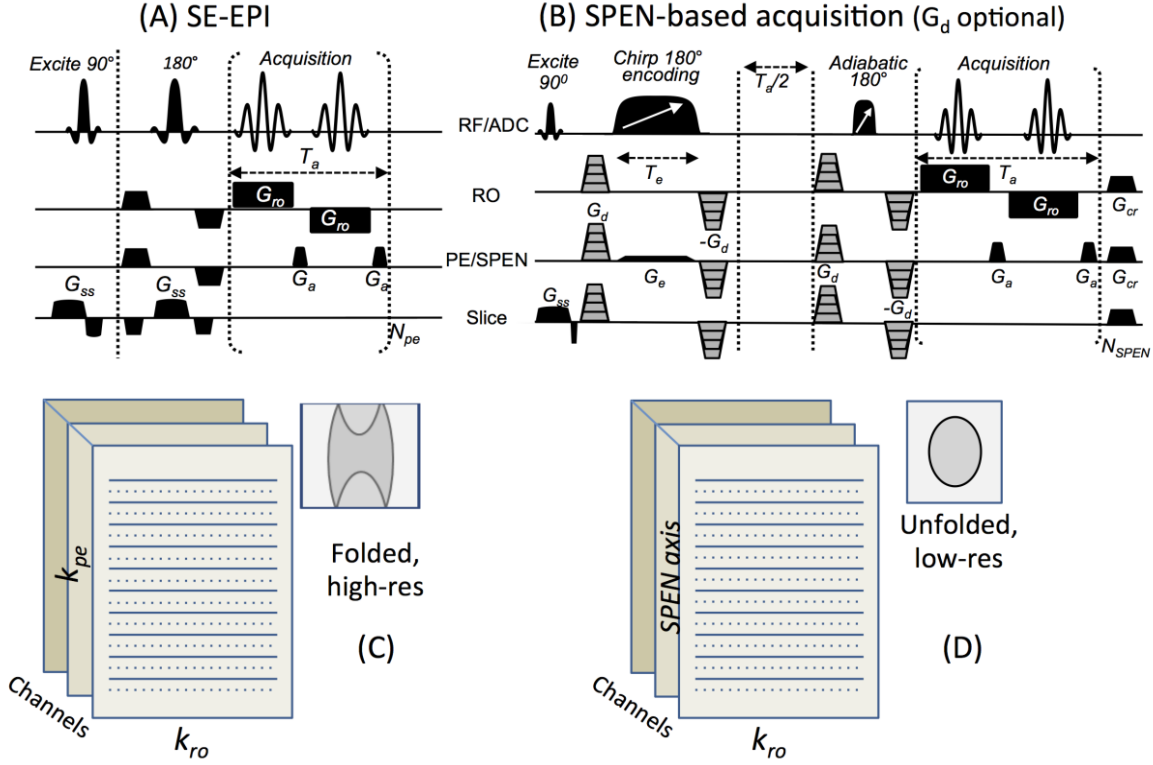

**Figure 1:** (A,B) Pulse sequences for SE-EPI and Hybrid SPEN. The latter involves an initial 90° slab-selective excitation, a pre-encoding $T_a/2$ delay inserted to achieve full-refocusing conditions, a 180° chirped encoding pulse, and a post-acquisition adiabatic 180° pulse returning all spins outside the targeted slice/slab back to thermal equilibrium (20). Optional doubly-refocused diffusion-weighting gradients $G_d$ in-between the 180° pulses are shown for ADC mapping; diagonal arrows indicate frequency modulations on the corresponding pulses. Additional definitions: RF/ADC: radiofrequency and analog-to-digital conversion channels; RO: read-out; cr: crusher; sp: spoiler. (C, D) Differing effects of skipping lines (dashed) in the low-bandwidth domains: whereas in EPI this leads to folding artifacts that require multiple sensors for unfolding and require either additional auto-calibration lines (30) or independent acquisitions to obtain the sensitivity maps, in SPEN they lead to lower-resolution but unfolded images, which can be improved by super-resolution procedures and used directly for sensitivity map calculations.

## METHODS

*The SUSPENSE algorithm.* As mentioned, the SPEN signal can be visualized as either a *k*-domain acquisition whose signal $S(k)$ is detected under the action of an acquisition gradient $G_a$, or as a direct image/spatial-domain acquisition over a time $T_a$ of the image being sought; i.e.,

$$S[k(t)] \propto \int_y \rho(y) e^{i[\varphi_e(y)+k_{pre}y+\gamma G_a y t]} dy = \int_y \rho(y) e^{i[\varphi_{total}(y)]} dy \approx \Delta y \cdot [\rho(y_{stat})] \left(\frac{\partial \varphi_{total}}{\partial y}\right)_{y=y_{stat}} = 0 \quad [1]$$

Here $k_{pre}$ is a suitable prephasing gradient that defines the beginning of the rasterization (-$\gamma G_a T_a/2$ in the case of encoding by an inversion pulse), and $y_{stat}$ is the coordinate that for a given $k(t) = \gamma G_a t$ ($0 \leq t \leq T_a$), fulfills the stationary phase condition. Assuming that signals are monitored over a suitable time $T_a$ (usually $2G_e T_e/G_a$ for an inversion-based encoding), the [$\partial\varphi/\partial y$] = 0 condition implies that $|S(k)|$ will be proportional to the $\rho(y)$ profile at uniformly spaced locations $y(k) = -\frac{FOV}{2} + \frac{FOV}{\gamma G_a T_a}k$. Extension of these arguments to a 2D hybrid SPEN acquisition requires retaining a discretized SPEN version of Eq. (1), which assuming the digitization of $1 \leq m \leq M$ points leads to

$$S[k_m] \propto \sum_n \rho(y_n) e^{i[\varphi_e(y_n) + k_{pre}y_n + k_m y_n]} \quad .\qquad [2]$$

where $1 \leq n \leq N$ is an index denoting the center of the $y_n$-th voxel. As in EPI, the sampled values of $k$ will usually be equispaced, $k_m = (m-1)\Delta k_{SPEN}$; unlike in EPI, however, these $\Delta k_{SPEN}$ values will not be given by Nyquist criteria. Instead, they will be defined by the bandwidth $BW = \gamma G_e FOV$ of the swept pulse used to impart the spatiotemporal encoding, by the duration $T_e$ of this pulse, and by the number $M$ of sampled blips. The chirped bandwidth in SPEN is usually set so that the Hz/pixel in the final image is sufficiently large to overcome field inhomogeneity distortions –normally 2-5 times stronger than for the low-bandwidth axis in a comparable EPI acquisition (alternatively, should these factors be set equal, there would be little point in performing SPEN rather than EPI). This bandwidth times the encoding time $T_e$, also referred to as the pulse's time-bandwidth product $Q = BW \cdot T_e$, dictate together with the FOV the curvature of the SPEN encoding parabola: $\alpha_{enc} = -Q/FOV^2$. Q therefore defines the resolution that can be obtained from a simple magnitude calculation of the data. In addition, Q will define the number of elements that should ideally be acquired in order to resolve all the resolvable spatial elements in the SPEN image. Indeed, for the fully-refocused sequences to be here considered, where isochromats span a bandwidth $BW$ of frequencies, the ideal dwell time $\Delta t$ of a SPEN sampling that aims to capture all spatial features would have to fulfill $\Delta t = \frac{1}{2BW}$. If this sampling is implemented by $M_{ideal}$ points spread over an acquisition time $T_a$, this in turn means that $\frac{T_a}{M_{ideal}} = \frac{1}{2BW}$ or, in other words, $M_{ideal} = BW \cdot T_a = 2Q$. This represents the ideal sampling that one would like to fully capture the available resolution, and collecting this many points will in general be feasible for unconstrained, 1D acquisitions. Yet for 2D single-shot sequences of the kind introduced in Figure 1, where SPEN needs to include gradient oscillations encoding the orthogonal $k_{ro}$ readout domain, acquiring this many samples will not be generally possible unless restricting the range of experimental $k_{ro}$ values, employing exceedingly long echo times TE, or relying on inordinately rapid and intense $G_{ro}$ gradients. As mentioned, such dense sub-sampling can be recapitulated by the interleaving procedure introduced in Ref. (27). SPEN interleaving makes up for undersampling by applying a suitable set of blips that advance the overall spatial sampling carried over $N_{shot}$ independent scans, in increments $\Delta k_l = (l-1)\frac{\Delta k_{SPEN}}{N_{shot}}$, $l = 1…N_{shot}$. In a suitably interleaved experiment, the numbers of sampled points $M_{spen}$ times $N_{shot}$ will then equal $2Q$.

Remarkably, the availability of multiple receivers sampling different regions throughout the targeted FOV, provide the means to compute the signals that would arise interleaved SPEN scans *in a single signal acquisition*. To visualize how this is feasible we recall that according to the SMASH formalism (29) spatial phase variations can, if sufficiently slow, be mimicked by summing signals arising from multiple coils: $\exp[-ih\Delta ky] = \sum_c n_c^h(y)S_c(y)$, where $n_c^h(y)$ are suitable weighting coefficients for the $h^{th}$ harmonic, and $S_c(y)$ are sensitivity maps for the various *c*-coils. Such formalism is usually of limited usefulness since the weighting coefficients need to hold throughout the entire $FOV_y$ (Figure 2A); in SPEN by contrast, where signals are emitted from localized spatial neighborhoods, it is simple and robust to extend such formalism for the sake of computing the "missing" interleaved data in a localized manner. To do so we look for a set of localized coefficients $n_c^h(x',y')$ such that the coils will satisfy, for a particular (*x',y'*)-neighborhood,

$$\sum_{c=1}^{N_c} n_c^h(x',y')W(y)S_c(x,y) = W(y)\exp\left[-ih\Delta k_{SPEN} \cdot y/N_{shot}\right], \qquad y \in FOV \qquad [3]$$

where $N_c$ is the number of coils and $W(y)$ is a weighting function which emphasizes the local nature of the SPEN interpolation. For this function we used a Gaussian, centered at the y' position being considered and weighted by a $\sigma = 1.2$, as this was found the best tradeoff between resolution and artifacts (lower σ's led to smoother lower resolution images while higher σ's led to a higher resolution albeit enhanced artifacts). This local interpolation enables one to faithfully synthesize localized harmonics up to a degree $R \leq N_c$ (Fig. 2B), leading to a resolution enhancement factor that takes the role that $N_{shot}$ played in the interleaved acquisitions. This *R*-factor multiplies the effective number of points collected along the SPEN axis, increasing it from $2 \cdot M_{spen}$ to $2 \cdot R \cdot M_{spen}$. For a good parallel receiving setup it is sensible to increase this number up to $M_{ideal} = 2 \cdot Q$. At this limit the effective oversampling has covered the maximum range of frequencies *BW* contained by the targeted FOV; sampling a larger bandwidth (i.e., interpolating further in *k*-space) will improve the nominal resolution, but neither enhance the actual spatial resolution nor lead to any additional improvements by SR algorithms. In fact at this full sampling limit, the SPEN experiment can be viewed as a conventional *k*-space acquisition on an object that has been imparted *a priori* with a quadratic phase. Thus the object's reconstruction can be carried out by a simple FT, without a need to carry out a sometimes ill-conditioned SR reconstruction.

An appealing aspect of this coil-based interpolation procedure is that –unlike what occurs when physically collecting interleaved SPEN scans or performing parallel *k*-based imaging– the value of R can be set and changed during a processing stage, until optimal images arise. In other words the role taken by this *R*-factor is entirely akin to that that is known from parallel imaging schemes –apart from the fact that its value can be chosen and optimized after completing, rather than before beginning, the acquisition. Another appealing difference between this SPEN resolution enhancement and *k*-space acquisitions relates to the fact that, like all parallel imaging approaches, SUSPENSE requires an *a priori* knowledge of the coil sensitivity maps $\{S_c(x,y)\}_{1 \leq c \leq N_C}$ to be used. The direct-space sampling nature of SPEN, however, implies that despite its low-bandwidth undersampling, all its pre-processed images are free from folding effects. The single-shot image to be interpolated therefore carries each coil's individual, unfolded sensitivity map; in order to extract these one can either calculate ratio images between each channel's signal (after smoothing) and the total root mean square (RMS) image, or rely on an algorithm like Eigenvalue-based Self-consistent Parallel Imaging Reconstruction (ESPIRiT, 33). In the present study we used both approaches and, in either case, the coil-mapping remained fully self-referenced and auto-calibrated.

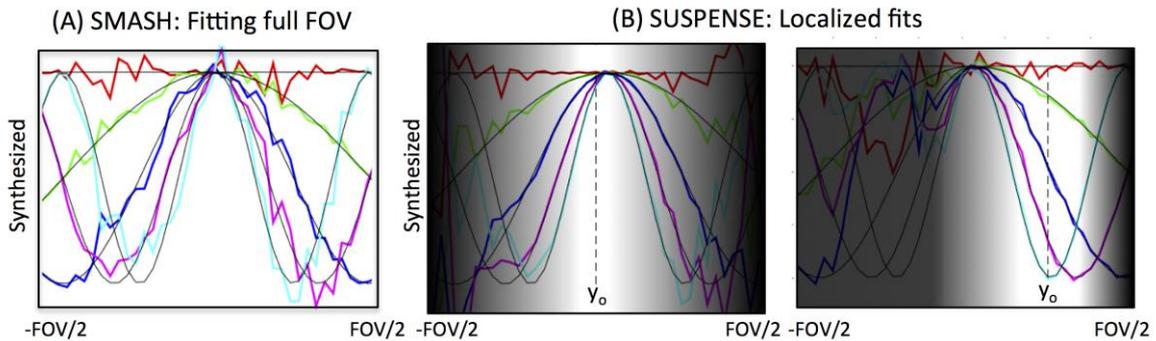

**Figure 2**: Comparison between (A) sensor-synthesized harmonics in conventional k-space imaging, and (B) locally accurate sensor-synthesized harmonics of the kind used in this study to interleave SPEN data. Different coefficients are used to syntesize a desired set of harmonics around the 2 different yo locations shown (brightest region in panels), with 0th-to-4th harmonics being shown with different colors.

While Eq. (3) implements the $k_{SPEN}$-space interpolation based on the SMASH formalism, it is feasible to carry out an essentially equivalent resolution enhancement procedure by SENSE-based reconstruction algorithms (28,31). In this case one would then search for an image $\rho$ satisfying $A\rho=S$, where *A* is an operator which includes a convolution by the quadratic phase that SPEN imparted on the spins, a weighting on the multiple sensors' sensitivities, as well as a Fourier transform (FT) implemented along the RO direction in the case of 2D hybrid acquisitions. Due to numerous missing SPEN/*k*-lines, searching for a high-definition $\rho$ image based on a direct inversion of $S=A\rho$, would lead to folding. Therefore an alternative approach was adopted whereby a high resolution, regularized image $\rho$ is sought that, after subject to a correct transformation described by the *A* operator, reproduces the measured signal. This *A* is an image-to-data transformation operator, so in this case it involved multiplication by the various coils' sensitivity maps, FT along the RO dimension and multiplication by a suitable SR matrix along the PE dimension (4,10) (for the multiband experiments, *A* also included summing over the contributions from the different bands). To guide this search in image space the non-linear conjugate gradient algorithm in (34) was adopted, using a sparse image representation based on a wavelet transformation and total-Variation (TV) regularization in the image search.

*Experimental methods*. Experiments were performed on a 3T Siemens TrioTIM (Siemens Healthcare, Germany) scanner using a 32-channel head-only coil. Phantom experiments to ascertain the resolution enhancement were run on a Bruker stripped phantom (12cm in diameter; ~0.8mm between stripes). Brain imaging scans were run on healthy volunteers in a protocol that included the acquisition of Turbo-Spin-Echo (TSE) reference images, SE-EPI images (Fig. 1A), and multislice SUSPENSE scans using the sequence in Figure 1B (with/without the diffusion gradient weighting, as determined by the experiment). This sequence was implemented using doubly-refocused fully-refocused formats that have already been described (19,20,35), but it included a number of innovations to increase volume coverage that are worth remarking. One of these involved the use of stimulated echoes capable of delivering numerous slices for each encoded slab (20,35). Another modification included the use of multiband pulses addressing different regions along the slice-selection ($z$) axis, chosen sufficiently distant to have different sensors in charge of their bulk detection (39). These multiband pulses were written as sums of simple *sincs* with Hamming windowing, and were 2.5-5ms long; no provisions were taken to optimize the SAR of these pulses. Notice that the parallel-receive axis exploited by these pulses lies perpendicular to the one discussed in the previous paragraph in connection to SUSPENSE, and to its $\Delta k_{SPEN}$ interpolation along *y*. Still, in such simultaneous multi-slice experiments, an additional constraint was added to the system of equations in Eq. 3, requiring that for each band the contribution of all other bands be zero. Mathematically this was cast as

$$\sum_{c=1}^{N_c} n_{c,m}^h(x',y') W(y) S_c^{m'}(x,y) = 0, \qquad y \in FOV, m \neq m' \qquad [4]$$

where $S_c^m(x,y)$ is the sensitivity map for the $m^{th}$ slice, and m' encompasses all other slices that were simultaneously excited together with one being processed. Finally, in order to facilitate the simultaneous acquisition of multiple nearby slices in a single scan, an approach analogous to EPI's simultaneous image-refocused (SIR, 36,37) was also implemented. In this approach a "$k_{ro}$-kick" is applied in-between subsequent excitation pulses, leading to multiple slice-specific echoes being resolved during each oscillating readout segment. As SAR in multi-slice SPEN stems predominantly from the application of the swept 180˚ encoding pulses in the sequence, all these procedures enabled us to widen the volume covered, without a concomitant increase in SAR. Typical SUSPENSE acquisitions thus managed to cover 20x18x7.2cm volumetric FOVs at 1x0.9x3mm spatial resolutions, with TE ≈ 40-50 ms in TR = 3sec. In all these volumetric studies SAR values ranged between 60% and 90% of the scanner's maximum prescribed values for brain analyses. Diffusion Weighted Imaging (DWI) maps were also run on volunteers by adding suitable sensitizing gradients (25,38), and compared against EPI scans collected using a scanner-supplied twice-refocused SE-EPI sequence (34). All studies were approved by the Institutional Review Board of the Wolfson Medical Center (Holon, Israel), and signed informed consents were obtained from all of the participants.

*Data processing*. To implement the SUSPENSE image reconstruction, in-house algorithms were written on Matlab® (Mathworks, Natick, MA), while all other images were processed on the scanner. When relying on SIR to encode multiple slabs along the slice-selection axis, the data for each echo was first separated using a simple splicing along the $k_{ro}$-space. In single-band protocols, sensitivity maps could be extracted directly from the same data set as used for the actual imaging: for this, standard SR was used to extract a well-resolved image per channel, from which sensitivity maps were calculated using either the ESPiRiT algorithm (33) or from the ratio between the smoothed images arising from each channel and the overall RMS image. In multiband runs this self-referenced procedure was not suitable, and sensitivity maps had to be obtained from separate acquisitions. These separate acquisitions employed identical parameters as the subsequent SPEN runs, apart from the use of conventional rather than multiband pulses; in order to make up for this change the number of collected slices was naturally increased, and with it increased the overall TR. With these sensitivity maps, two approaches were developed to reconstruct the final high-resolution image –both yielding results of similar qualities. In the first one, the missing SPEN lines were calculated by locally combining the data from the different sensors (Eq. (3)) calculating the required coefficients for the higher-order harmonics using a Moore-Penrose inverse of the constraints matrix with a small Tikhonov regularization ($\lambda$=0.001). These coefficients were calculated for each (x,y,z) location (see Fig. 2 for calculations for particular ($x_o,z_o$)-coordinate), and *g*-factor maps (28) were calculated for each location and each harmonic. With the aid of these harmonics the missing lines in the SPEN data set were reconstructed, and the final image was calculated by applying the super-resolution algorithm described in (22) on the resulting augmented set. In an alternative rendering of this processing, an image-based SENSE reconstruction procedure was implemented, using the nonlinear conjugate gradient algorithm in Ref. (34) with TV regularization. This algorithm required implementing image-to-data transformations (and their conjugate transform), which for the hybrid SPEN acquisitions required (a) the application of an appropriate parabolic phase, (b) a FT along the RO dimension, and (c) suitably accounting for the different channels sensitivities. The code for the ensuing SUSPENSE algorithm is available upon request. For the diffusion experiments, Apparent Diffusion Coefficient (ADC) maps were calculated voxel-wise from data measured using a $b_0 = 0$ and three $b = 650$s/mm$^2$ values achieved by orthogonal diffusion-weighting gradients. Additionally, when multiple scans where averaged to improve the sensitivity of diffusion measurements, an $L_2$ regularization was used in order enable the averaging of the single-shot SPEN data in image space; this averaging was rendered free from phase complications and instabilities, by performing it either in magnitude or after a low-resolution phase correction of the 2D images.

**RESULTS**

Figure 3 summarizes the main steps used in the resolution enhancing procedure hereby introduced, using actual single-slice brain data as illustration. The procedure begins by resolving the various slices and FT of the data along the $k_{ro}$ axis, to yield for a given slice a set of low-resolution 2D images per sensor. The quality of these raw images is improved using a super-resolution algorithm (Fig. 4, left-hand column), and from these improved images the sensitivity maps $\{S_c(x,y)\}_{1\leq c\leq N_c}$ are obtained as detailed in Methods. Solving Eq. 3 for each location allows one to obtain the desired spatially-dependent coefficient maps, as illustrated in the lower-left panel of Fig. 3 for $h = 2$. These coefficients are then used to complete the missing data lines (Fig. 3, center-right panel), and a super-resolution procedure on the resulting set involving a deconvolution with the quadratic phase kernel, yields the high-definition image being sought. Notice that this final full image (Fig 3, lower-right) is obtained by implementing an RMS combination of the different sensor's data yielding, as happens in the case of GRAPPA (31), a concurrent improvement in SNR.

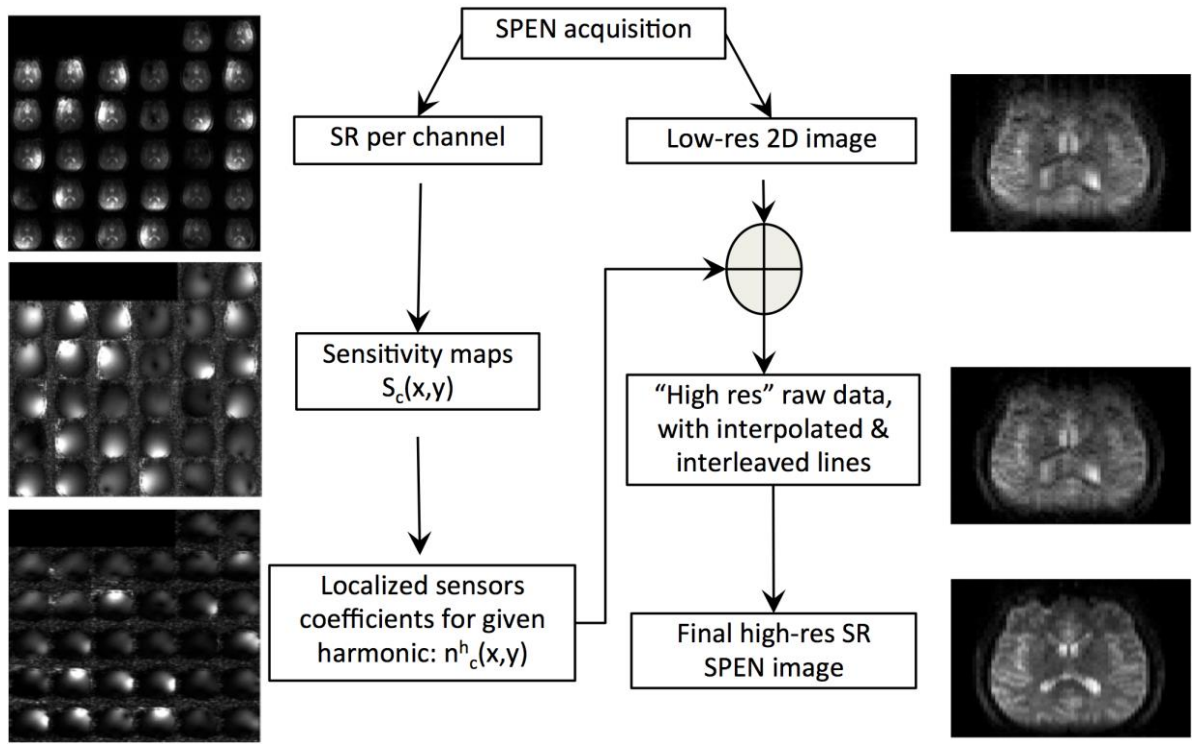

**Figure 3**: Basic algorithm for interpolating single-shot SPEN data in its k/image space based on manipulations that synthesize the missing data from higher-order coil harmonics. The left-hand side describes the procedure used for obtaining the coefficients maps for a given spatial harmonics, based on channel-per-channel SR-enhanced images leading the sensitivity- and coefficient-maps being sought. The right-hand column shows how these coefficients permitted a reconstruction of the missing lines, leading to a substantial resolution enhancement at no extra experimental cost. See the main text for an alternative, image-based reconstruction algorithm.

Figure 4 shows the effects of this procedure on single-shot SPEN results collected on a phantom, where the SPEN axis is placed along the vertical dimension. This phantom possesses a number of carved features, including a stripped structure with slots spaced by ~0.8mm. While all the images shown were obtained from the same single-scan data set and processed using super-resolution, their definitions along the SPEN axis are clearly different.

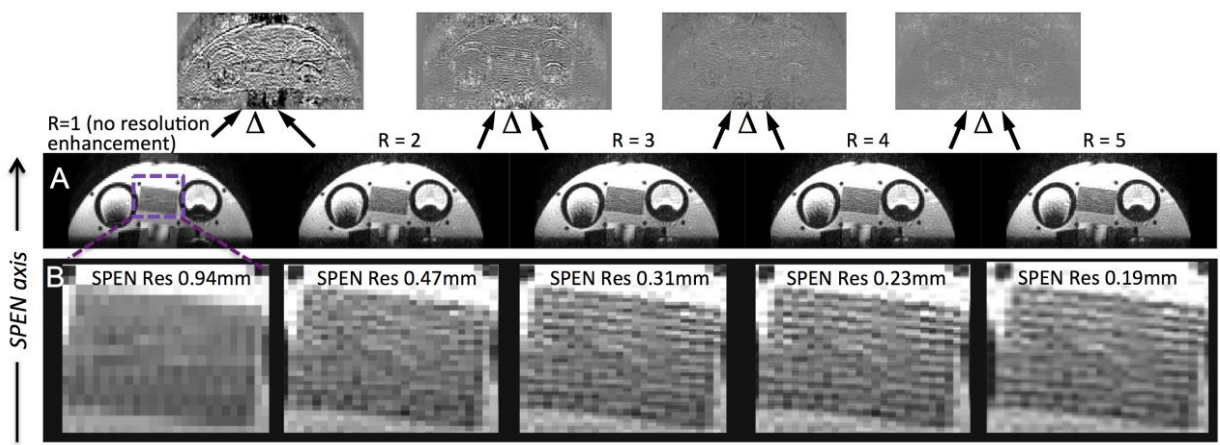

**Figure 4**: Resolution-enhanced processing of phantom-based SPEN data. (A) Single-shot phantom images processed with different reconstruction enhancements as summarized in Figure 3, using R factors ranging from 1 to 5. Basic acquisition parameters: FOV = 15x4.5cm, SPEN Q = 65, Ge = 0.1G/cm, Te = 31.7ms and Mspen = 48, in-plane matrix size after SR = 200x48, i.e. 0.75x0.9375mm resolution. (B) Zoomed-in displays showing the improvement brought about by the processing, onto a high-spatial-frequency portion of the phantom, as the nominal SPEN axis resolution drops well below 1mm. Shown on top of the figure are difference ($\Delta$) images between consecutive renderings of the data, evidencing the absence of changes/improvements past R ≈ 3.

This can be appreciated by comparing the slotted blow-outs shown on the lower row of the Figure. Notice that although additional *k*-rows can be interpolated up to the number of available receiver coils, the images show no resolution improvements past a R ≈ 3 factor. This can be appreciated in the images as well as by the differences calculated between consecutive interpolation steps, shown on the top row of Figure 3. This is in agreement with the above-mentioned expectations, whereby extending the oversampling beyond 2Q (130 for this phantom) does not lead to an actual spatial resolution improvement but simply to a data interpolation.

Figure 5 illustrates further the details of this procedure, using a volunteer brain scan as example. Stimulated-echo SPEN allowed us to cover a full brain FOV of 20x18x7.2cm with 24 slices in 3sec (for better appreciating the details of the processing, these images are shown after signal averaging 80 identical scans in image space). The original resolution of the SPEN acquisition was set to 1x4.5x3mm along the RO/SPEN/SS dimensions, using $M_{spen}$ = 40 samples. Figure 5A shows a representative slice arising after super-resolution of these data without SUSPENSE; Figure 5B shows the same slice after SUSPENSE is implemented with R = 5, leading to the effective sampling of all 2Q = 200 values associated to the acquisition. The resolution improvement along the SPEN axis is evident, even if there is a noticeable "stripping" when comparing it against a multiscan TSE counterpart collected at a similar 1x1x3mm resolution. We have traced this artifact to residual even/odd effects that remain to be corrected in the SUSPENSE acquisition. Figures 5D-5E present additional aspects of this multi-coil processing, including the sensitivity maps $\{S_c(x,y)\}_{1 \leq c \leq 32}$ (magnitude and phase) afforded by ESPIRiT from the original SPEN images, and the $\{n_c^{h=2}(x,y)\}_{1 \leq c \leq 32}$ spatial coefficients obtained from the sensitivity maps for the 2$^{nd}$ spherical harmonic set required for implementing R = 5 SUSPENSE (only the first 5 channels shown for clarity). From the first of these data sets g-factor maps can be calculated, describing the degree of independence of the various sensors. Noise amplification factors were calculated using a multiple-replica approach whereby *a priori* known synthetic noise was artificially fed throughout the targeted FOV, and its value at the conclusion of the SUSPENSE processing was evaluated. Figure 5F illustrate the ensuing noise maps, for R = 5 interpolations executed using the SENSE algorithm.

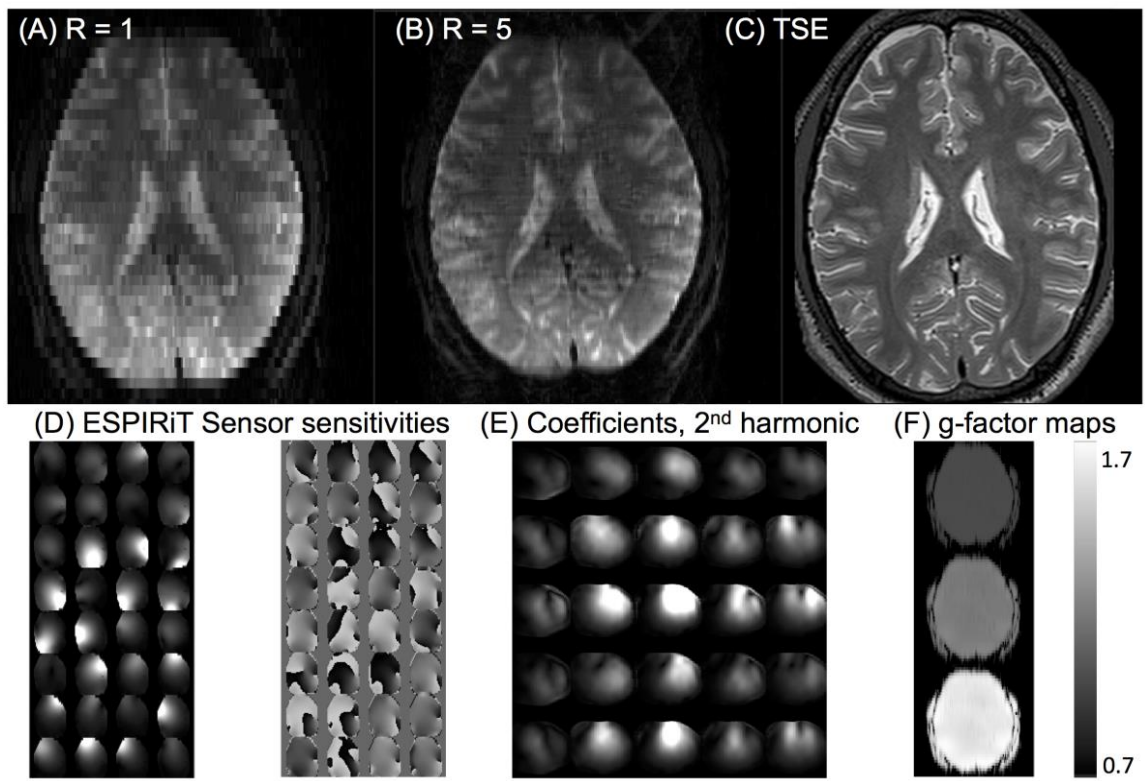

**Figure 5**: (A,B) Idem as in Fig. 4 but for acquisitions performed on a human volunteer, for an in-plane FOV=20x18cm. The initial resolution=1x4.5mm was improved to 1x0.9mm with an R=5 factor, chosen to fulfill Mspen.R=2Q=200 (as Te=20ms, BW=5kHz). For better evidencing SUSPENSE's resolution improvements these images arise from averaging 80 identical, separately-processed repetitions. (C) TSE anatomical image collected at 1x1mm in-plane resolution. Additional aspects of this processing approach include magnitude and phase sensitivity maps derived from ESPIRiT (D), some of the coefficient maps calculated for h=2 (E), and a noise amplification map calculated for R=5 SUSPENSE (F). The latter estimation was calculated using a synthetic-noise multiple-replica approach, followed by SUSPENSE reconstruction based on the SENSE algorithm with a linear, norm-2 regularization. Minor ghosting artifacts visible in the R = 5 image reflect departures from the stationary phase approximation (8).

As mentioned in the theoretical considerations, as the number of sensor-interpolated points increases, the gains that can be conveyed by subjecting SPEN data to super-resolution vanish. Figure 6 illustrates this by comparing, using the same data set as introduced in Figure 5, the results that can be retrieved for R = 15, upon applying the SR vs a conventional FT along the interpolated axis. The virtual identity between the two images reflects the fact that, for a sufficiently dense and accurate interpolation, all the frequency range that was involved in the SPEN acquisition –as given by the bandwidth of the FOV that was encoded by the chirp pulse– becomes faithfully characterized by SUSPENSE's oversampling. This shows that given a sufficient interpolation all complications associated to SPEN's post-processing methods can be bypassed without relinquishing on SPEN's immunity to inhomogeneities.

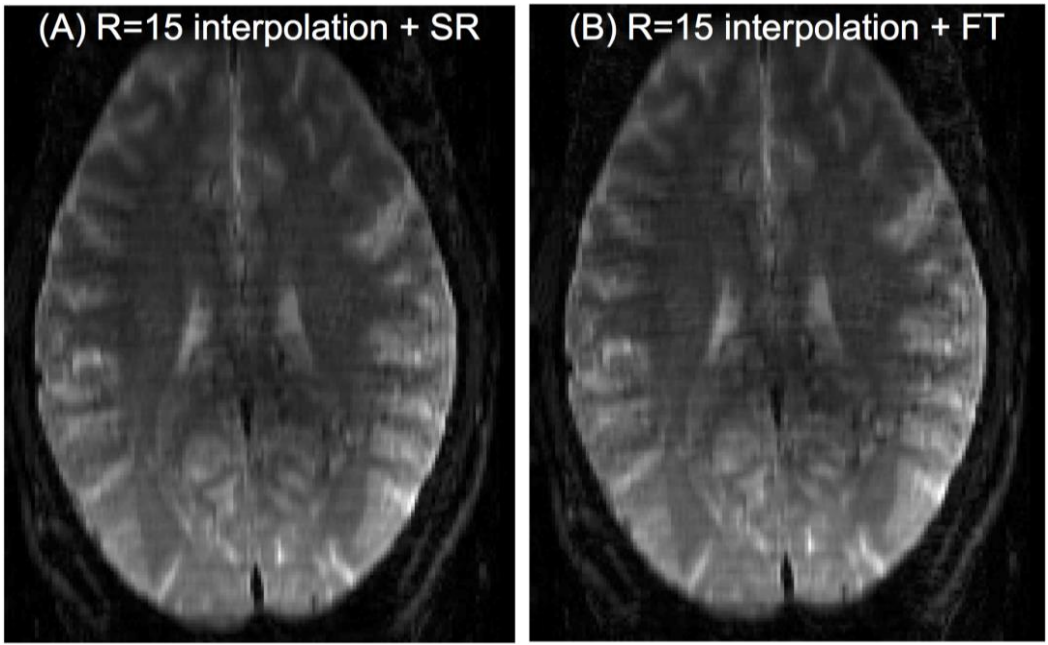

**Figure 6**: Consequences of a very high k-space SPEN interpolation: as the spatial voxel size upon using a high R-value (in the present case 15) becomes very small, the intra-voxel dephasing due to the quadratic phase weakens. Hence one obtains a similar result by approximating the image $\rho$ using the super-resolution procedure of Fig. 2, or by a simple $\rho \approx$ FT[S(kSUSPENSE)] transform (the resulting image then contains a quadratic phase, not seen in these magnitude-mode displays).

Figure 7 illustrates the SENSE-based interpolation algorithm, using a diffusivity measurement as test. To implement the latter, pairs of doubly-refocused diffusion gradients were applied along orthogonal axes in different scans (Fig. 7A). These results illustrate an attractive feature of the ensuing approach: given SUSPENSE's robustness this procedure can be implemented on a scan-by-scan basis, and the resulting images co-added in magnitude mode without suffering from phase inconsistencies. This in turn endows excellent sensitivity to the final diffusion maps, despite their sub-mm resolution. Comparisons against conventional EPI-derived ADC maps (Fig. 7B) clearly evidence these resolution improvements, while certifying the correctness of the SUSPENSE diffusion maps when compared against single-shot counterparts. Additional data provided in the Supplementary Information (Supporting Figures S1 and S2) demonstrates the possibility of retrieving this kind of high definition data with a rapid and sizable volumetric coverage, while retaining low SAR values.

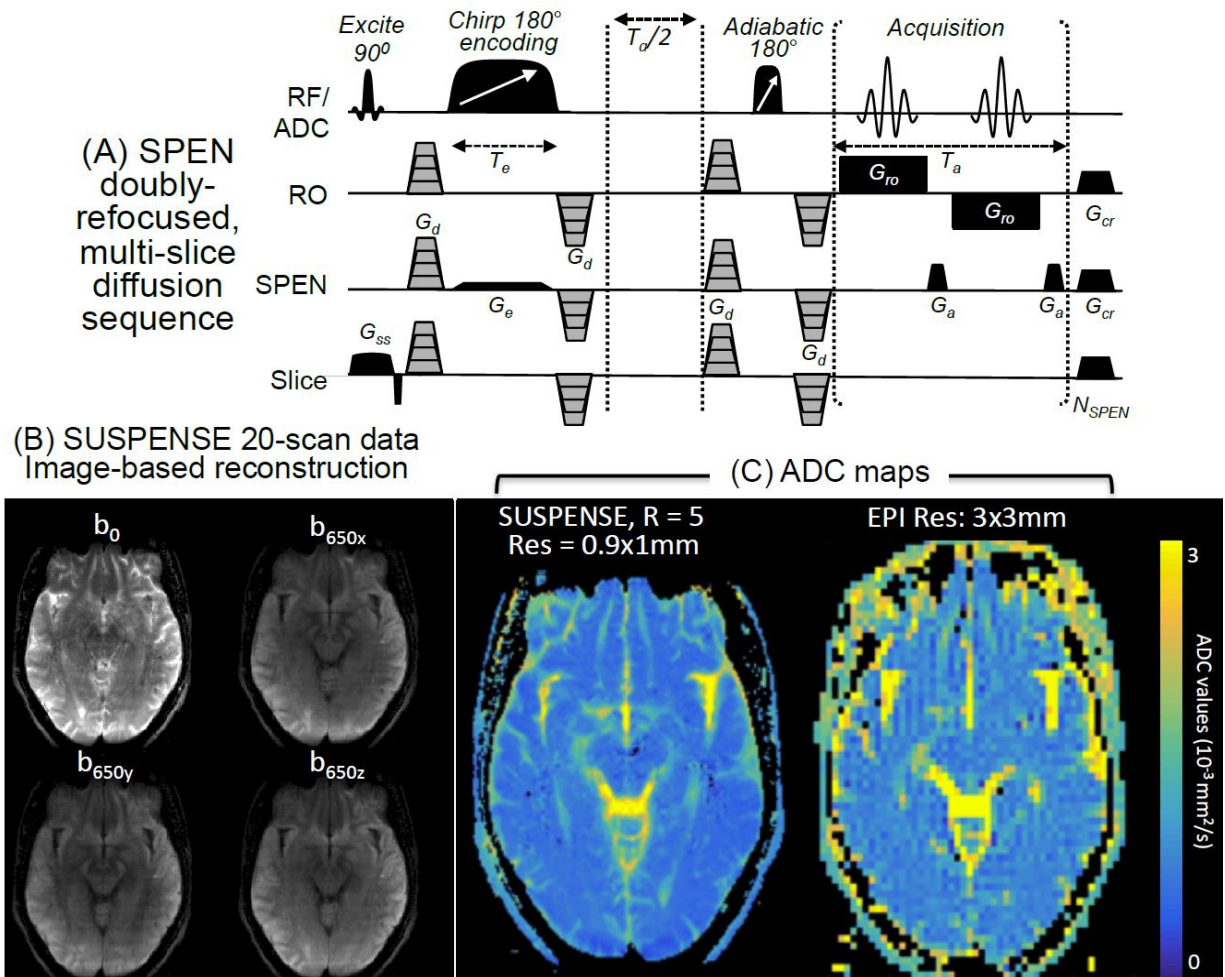

**Figure 7**: Diffusion weighted brain images illustrating the quality achievable by combining the SUSPENSE single-shot sensor-based interpolation with multi-scan averaging. (A) R = 5 SUSPENSE b-weighted images arising after 20 averages, collected using the indicated nominal b-values. Images (13 slices) were originally acquired with a 20x20x10.4cm FOV and a 4.5x1x2mm resolution. Total experimental time = 14min. (B) Comparison between the ADC maps arising from the interpolated SUSPENSE vs EPI experiments collected on the same FOV –the latter with a 2x2x2mm spatial resolution. Notice that despite the uneven T2 contrast evidenced by the SPEN images owing to the sequential image unraveling along the anterior to posterior direction, this effect factors out upon computing a normalized contrast like diffusivity. In this case SPEN images were processed using a SENSE-based reconstruction. The EPI map arises from a four-scan signal averaged acquisition lasting ca. 1min.

## DISCUSSION AND CONCLUSIONS

This study discussed a new approach for improving the resolution of hybrid SPEN acquisitions, at no expense in experimental complexity or increase in acquisition scans. This approach stems from the realization that in SPEN the acquisition wavevector plays a dual role –delivering at the same time the image in direct space, while sampling a reciprocal space. Image interleaving procedures that had been hitherto performed in multiple scans, could therefore be replaced by *k*-space interpolation procedures based on multiple receivers. This delivered higher resolution images at no additional expense in the experimental protocol, using information that had hitherto not been exploited in this kind of acquisitions. The approach is certainly not general as it requires a sufficiently high number of independent

sensors –a commodity that is rarely available in preclinical settings, and not always efficiently implemented in human scanners. When parallel imaging facilities are available, however, SUSPENSE's images exhibit substantially better resolution than their SPEN counterparts. Furthermore, with the highest g-factors observed for the highest harmonics remaining ≤3, reasonable SNR losses are associated with this increased resolution. Another attractive feature is the method's self-referenced, auto-calibrated nature. This full self-reliance opens the possibility of averaging signals using magnitude-mode, direct-space single-shot images – even when involving diffusion weighting gradients. Unlike previously reported multi-shot interleaved EPI or SPEN acquisitions, this opens the possibility of reaching both high-resolution and high-sensitivity ADC maps, as illustrated in Figure 5. A third notable feature of SUSPENSE is the possibility of reconstructing its higher-definition images in a variety of ways. In the present case we demonstrated approaches based on the synthesis of higher-order local *k*-harmonics or on image-domain reconstruction procedures; these procedures parallel concepts underlying SMASH and SENSE respectively, yet it is conceivable that additional approaches can also be devised. A final point worth remarking is the method's compatibility with all other improvements that have been hitherto developed for SPEN-like sequences, as demonstrated with the incorporation of multi-band, SIR or stimulated-echo approaches (Supplementary Information). A similar opportunity arises vis-à-vis improving the resolution of other variants like bi-SPEN or xSPEN (40), as will be discussed in upcoming studies.

**Acknowledgments.** We are grateful to Dr. Sagit Shushan (Wolfson Medical Center) and the Weizmann MRI team (Edna Furman-Haran, Fanny Attar and Nachum Stern) for assistance in the human scans. Financial support from the Israel Science Foundation grant 795/13, the EU through ERC-2016-PoC grant # 751106, Minerva funding (#712277) from the Federal German Ministry for Education and Research, the Kimmel Institute for Magnetic Resonance and the generosity of the Perlman Family Foundation, are acknowledged. ML also acknowledges the Weizmann Institute for a Visiting Faculty Program Fellowship.

**Figure Captions**

**Figure 1:** (A,B) Pulse sequences for SE-EPI and Hybrid SPEN. The latter involves an initial 90° slab-selective excitation, a pre-encoding $T_a/2$ delay inserted to achieve full-refocusing conditions, a 180° chirped encoding pulse, and a post-acquisition adiabatic 180° pulse returning all spins outside the targeted slice/slab back to thermal equilibrium (20). Optional doubly-refocused diffusion-weighting gradients $G_d$ in-between the 180° pulses are shown for ADC mapping; diagonal arrows indicate frequency modulations on the corresponding pulses. Additional definitions: RF/ADC: radiofrequency and analog-to-digital conversion channels; RO: read-out; cr: crusher; sp: spoiler. (C, D) Differing effects of skipping lines (dashed) in the low-bandwidth domains: whereas in EPI this leads to folding artifacts that require multiple sensors for unfolding and require either additional auto-calibration lines (30) or independent acquisitions to obtain the sensitivity maps, in SPEN they lead to lower-resolution but unfolded images, which can be improved by super-resolution procedures and used directly for sensitivity map calculations.

**Figure 2:** Comparison between (A) sensor-synthesized harmonics in conventional *k*-space imaging, and (B) locally accurate sensor-synthesized harmonics of the kind used in this study to interleave SPEN data. Different coefficients are used to syntesize a desired set of harmonics around the 2 different $y_o$ locations shown (brightest region in panels), with $0^{th}$-to-$4^{th}$ harmonics being shown with different colors.

**Figure 3:** Basic algorithm for interpolating single-shot SPEN data in its *k*/image space based on manipulations that synthesize the missing data from higher-order coil harmonics. The left-hand side describes the procedure used for obtaining the coefficients maps for a given spatial harmonics, based on channel-per-channel SR-enhanced images leading the sensitivity- and coefficient-maps being sought. The right-hand column shows how these coefficients permitted a reconstruction of the missing lines, leading to a substantial resolution enhancement at no extra experimental cost. See the main text for an alternative, image-based reconstruction algorithm.

**Figure 4:** Resolution-enhanced processing of phantom-based SPEN data. (A) Single-shot phantom images processed with different reconstruction enhancements as summarized in Figure 3, using R factors ranging from 1 to 5. Basic acquisition parameters: FOV = 15x4.5cm, SPEN $Q = 65$, $G_e = 0.1$G/cm, $T_e = 31.7$ms and $M_{spen} = 48$, in-plane matrix size after SR = 200x48, i.e. 0.75x0.9375mm resolution. (B) Zoomed-in displays showing the improvement brought about by the processing, onto a high-spatial-frequency portion of the phantom, as the nominal SPEN axis resolution drops well below 1mm. Shown on top of the figure are difference ($\Delta$) images between consecutive renderings of the data, evidencing the absence of changes/improvements past R ≈ 3.

**Figure 5:** (A,B) Idem as in Fig. 4 but for acquisitions performed on a human volunteer, for an in-plane FOV=20x18cm. The initial resolution=1x4.5mm was improved to 1x0.9mm with an R=5 factor, chosen to fulfill $M_{spen} \cdot R = 2Q = 200$ (as $T_e$=20ms, BW=5kHz). For better evidencing SUSPENSE's resolution improvements these images arise from averaging 80 identical, separately-processed repetitions. (C) TSE anatomical image collected at 1x1mm in-plane

resolution. Additional aspects of this processing approach include magnitude and phase sensitivity maps derived from ESPIRiT (D), some of the coefficient maps calculated for *h*=2 (E), and a noise amplification map calculated for R=5 SUSPENSE (F). The latter estimation was calculated using a synthetic-noise multiple-replica approach, followed by SUSPENSE reconstruction based on the SENSE algorithm with a linear, norm-2 regularization. Minor ghosting artifacts visible in the R = 5 image reflect departures from the stationary phase approximation (8).

**Figure 6:** Consequences of a very high *k*-space SPEN interpolation: as the spatial voxel size upon using a high R-value (in the present case 15) becomes very small, the intra-voxel dephasing due to the quadratic phase weakens. Hence one obtains a similar result by approximating the image $\rho$ using the super-resolution procedure of Fig. 2, or by a simple $\rho \approx FT[S(k_{SUSPENSE})]$ transform (the resulting image then contains a quadratic phase, not seen in these magnitude-mode displays).

**Figure 7:** Diffusion weighted brain images illustrating the quality achievable by combining the SUSPENSE single-shot sensor-based interpolation with multi-scan averaging. (A) R = 5 SUSPENSE *b*-weighted images arising after 20 averages, collected using the indicated nominal *b*-values. Images (13 slices) were originally acquired with a 20x20x10.4cm FOV and a 4.5x1x2mm resolution. Total experimental time = 14min. (B) Comparison between the ADC maps arising from the interpolated SUSPENSE vs EPI experiments collected on the same FOV –the latter with a 2x2x2mm spatial resolution. Notice that despite the uneven $T_2$ contrast evidenced by the SPEN images owing to the sequential image unraveling along the anterior to posterior direction, this effect factors out upon computing a normalized contrast like diffusivity. In this case SPEN images were processed using a SENSE-based reconstruction. The EPI map arises from a four-scan signal averaged acquisition lasting ca. 1min.

**Supplementary Figure S1:** (A) Extending the coverage afforded by SUSPENSE by means of multi-band excitation along the slab-selective axis (two bands), plus SIR-based slice deconvolution (colored pulses within each band and colored FIDs; detailed processing not shown) within the sensor-resolved multi-band slabs. The sequence shows an option stimulated echo block enabling an even higher multi-slicing without SAR penalties (3), which was not used in this example. (B) 2D SPEN images following separation of the two slice-selective echoes along $G_{RO}$ and FT along this axis, showing the multi-band overlap. (B) SENSE-based separation of the multibands, leading after SR to images with 1x2x2mm resolution and to the (x,y)-coil sensitivity maps. (C) SUSPENSE processing of the same data using a second-order spherical harmonic, leading to 1x1x2mm resolutions. Lower insets show

identical (B,C) slices, zoomed to highlight the finer delineation evidenced by the gray matter sulci (contoured in red). Additional acquisition parameters included the acquisition of 16 slices from a 20x8.4x2.6cm FOV in 6secs, using $G_e$ = 0.13G/cm, $T_e$ = 43.7ms and $M_{spen}$ = 42.

**Supplementary Figure S2:** Extending the coverage afforded by SUSPENSE by combining, along the slice-selective axis, a two-band excitation with the acquisition of six stimulated echo images arising from each encoded slab (2,6). Full brain coverage with 48 slices (4 slabs x 2 bands x 6 stimulated-echoes/slab) was thus achieved in a TR = 3sec. Images were processed with SUSPENSE and R = 5, leading to 1x0.9x2.5mm resolutions over a 20x18x12cm FOV. Green rectangles exemplify data that was simultaneously encoded in two bands. The images within each rectangle arise from different stimulated-echoes; i.e., to sequentially excited slices encoded within the same slab.

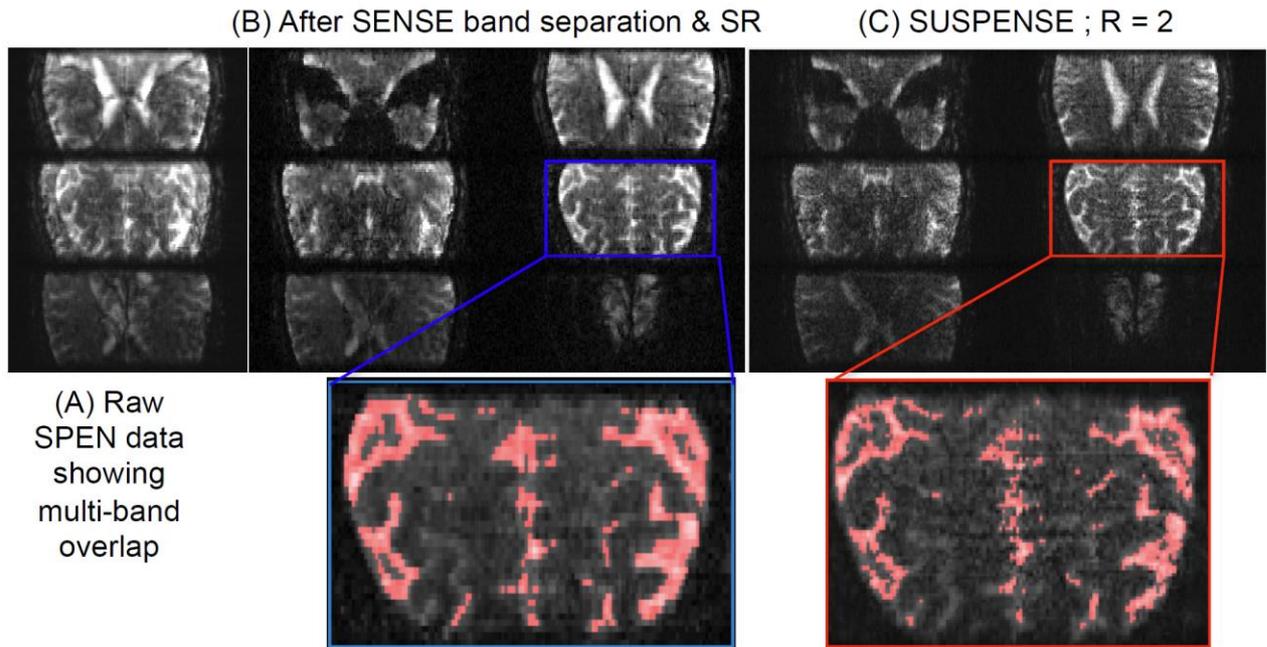

**Figure S1**: (A) Extending the coverage afforded by SUSPENSE by means of multi-band excitation along the slab-selective axis (two bands), plus SIR-based slice deconvolution (colored pulses within each band and colored FIDs; detailed processing not shown) within the sensor-resolved multi-band slabs. The sequence shows an option stimulated echo block enabling an even higher multi-slicing without SAR penalties (3), which was not used in this example. (B) 2D SPEN images following separation of the two slice-selective echoes along GRO and FT along this axis, showing the multi-band overlap. (B) SENSE-based separation of the multibands, leading after SR to images with 1x2x2mm resolution and to the (x,y)-coil sensitivity maps. (C) SUSPENSE processing of the same data using a second-order spherical harmonic, leading to 1x1x2mm resolutions. Lower insets show identical (B,C) slices, zoomed to highlight the finer delineation evidenced by the gray matter sulci (contoured in red). Additional acquisition parameters included the acquisition of 16 slices from a 20x8.4x2.6cm FOV in 6secs, using Ge = 0.13G/cm, Te = 43.7ms and Mspen = 42.

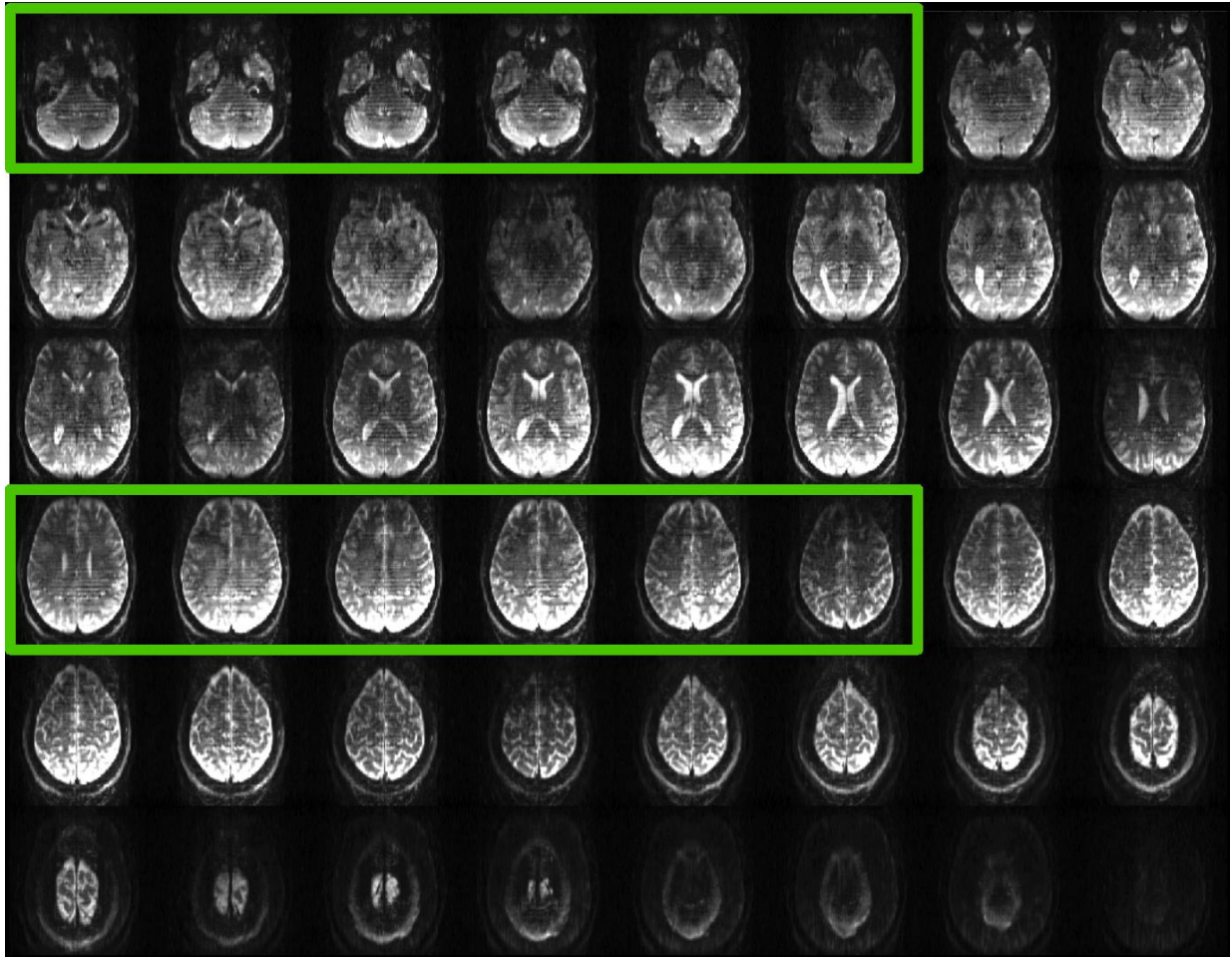

**Figure S2**: Extending the coverage afforded by SUSPENSE by combining, along the slice-selective axis, a two-band excitation with the acquisition of six stimulated echo images arising from each encoded slab (2,6). Full brain coverage with 48 slices (4 slabs x 2 bands x 6 stimulated-echoes/slab) was thus achieved in a TR = 3sec. Images were processed with SUSPENSE and R = 5, leading to 1x0.9x2.5mm resolutions over a 20x18x12cm FOV. Green rectangles exemplify data that was simultaneously encoded in two bands. The images within each rectangle arise from different stimulated-echoes; i.e., to sequentially excited slices encoded within the same slab.